\pdfoutput=1
\documentclass[sigconf]{acmart}

\AtBeginDocument{%
  }

\copyrightyear{2025}
\acmYear{2025}
\setcopyright{rightsretained}
\acmConference[CF '25]{22nd ACM International Conference on Computing Frontiers}{May 28--30, 2025}{Cagliari, Italy}
\acmBooktitle{22nd ACM International Conference on Computing Frontiers (CF '25), May 28--30, 2025, Cagliari, Italy}\acmDOI{10.1145/3719276.3725203}
\acmISBN{979-8-4007-1528-0/2025/05}

\acmConference[CF'25]{The 22nd ACM International Conference on Computing Frontiers}{May 28--30,
  2025}{Cagliari, Sardinia, Italy}
\acmISBN{979-8-4007-1528-0/2025/05}

\usepackage{gensymb}

\usepackage[ruled,vlined]{algorithm2e}
\SetKwInput{KwInput}{Input}              
\SetKwInput{KwOutput}{Output}
\SetKwInput{KwGlobal}{Global}

\usepackage{graphicx}

\begin{document}

\title[FractalSync]{FractalSync: Lightweight Scalable Global Synchronization of Massive Bulk Synchronous Parallel AI Accelerators}

\author{Victor Isachi}
\affiliation{%
  \institution{University of Bologna}
  \city{Bologna}
  \country{Italy}
}
\email{victor.isachi@unibo.it}

\author{Alessandro Nadalini}
\affiliation{%
  \institution{University of Bologna}
  \city{Bologna}
  \country{Italy}
}
\email{alessandro.nadalini3@unibo.it}

\author{Riccardo Fiorani Gallotta}
\affiliation{%
  \institution{University of Pavia}
  \city{Pavia}
  \country{Italy}
}
\affiliation{%
  \institution{University of Bologna}
  \city{Bologna}
  \country{Italy}
}
\email{riccardo.fiorani3@unibo.it}

\author{Angelo Garofalo}
\affiliation{%
  \institution{University of Bologna}
  \city{Bologna}
  \country{Italy}
}
\email{angelo.garofalo@unibo.it}

\author{Francesco Conti}
\affiliation{%
  \institution{University of Bologna}
  \city{Bologna}
  \country{Italy}
}
\email{f.conti@unibo.it}

\author{Davide Rossi}
\affiliation{%
  \institution{University of Bologna}
  \city{Bologna}
  \country{Italy}
}
\email{davide.rossi@unibo.it}

\renewcommand{\shortauthors}{Isachi et al.}

\begin{abstract}
  The slow-down of technology scaling and the emergence of Artificial Intelligence (AI) workloads have led computer architects to increasingly exploit parallelization coupled with hardware acceleration to keep pushing the performance envelope. However, this solution comes with the challenge of synchronization of processing elements (PEs) in massive heterogeneous many-core platforms. To address this challenge, we propose \textit{FractalSync}, a hardware accelerated synchronization mechanism for Bulk Synchronous Parallel (BSP) systems. We integrate \textit{FractalSync} in \textit{MAGIA}, a scalable tile-based AI accelerator, with each tile featuring a RISC-V-coupled matrix-multiplication (MatMul) accelerator, scratchpad memory (SPM), and a DMA connected to a global mesh Network-on-Chip (NoC). We study the scalability of the proposed barrier synchronization scheme on tile meshes ranging from 2$\times$2 PEs to 16$\times$16 PEs to evaluate its design boudaries. Compared to a synchronization scheme based on software atomic memory operations (AMOs), the proposed solution achieves up to $43\times$ speedup on synchronization, introducing a negligible area overhead ($<0.01\%$). \textit{FractalSync} closes timing at \textit{MAGIA}'s target 1GHz frequency.
\end{abstract}



\maketitle

\section{Introduction}
Massively parallel systems (MPSs) keep rising in popularity thanks to an increasing amount of emerging workloads, namely generative Artificial Intelligence (GenAI) and Large Language Models (LLMs), that rely on high-performance parallel systems. A common theme among modern Machine Learning (ML) accelerators is their exploitation of the intrinsic parallelism of workloads by relying on many processing elements (PEs). Often, these PEs need to synchronize periodically. As systems scale out, communication and synchronization overheads scale accordingly. Thus, for AI-oriented MPSs the presence of a fast and scalable synchronization mechanism is of crucial importance. 

The Bulk Synchronous Parallel (BSP) model~\cite{valiant1990bridging} is used to bridge the SW/HW gap in parallel systems. BSP allows us to more easily think about the hardware software interface by introducing guidelines on how they interact. This model is characterized by its popularity, ease-of-use, and generality \cite{gerbessiotis1994direct, cheatham1996bulk, Heng_McColl_2021}, and can be used for a variety of problems including ML \cite{Heng_McColl_2021}. Importantly, it relies on barrier synchronization as the only synchronization primitive.

In this work, we introduce \textit{FractalSync}, a hardware-accelerated barrier synchronization mechanism that is compatible with BSP and designed for modern GenAI architectures and MPSs. We integrate \textit{FractalSync} in \textit{MAGIA} as a proxy for a whole class of mesh-based MPSs.
\textit{MAGIA} features a 2D mesh of compute tiles connected to a global NoC; each tile is equipped with a dedicated General Matrix Multiplication (GeMM) accelerator, a DMA engine for data movement, and a large L1 SRAM. Finally, we study the performance gains and area overhead of our addition of \textit{FractalSync} into \textit{MAGIA}.

\section{MAGIA} \label{sec:magia}

\subsection{Tile}
\begin{figure}[t]
  \centering
  \includegraphics[width=\linewidth]{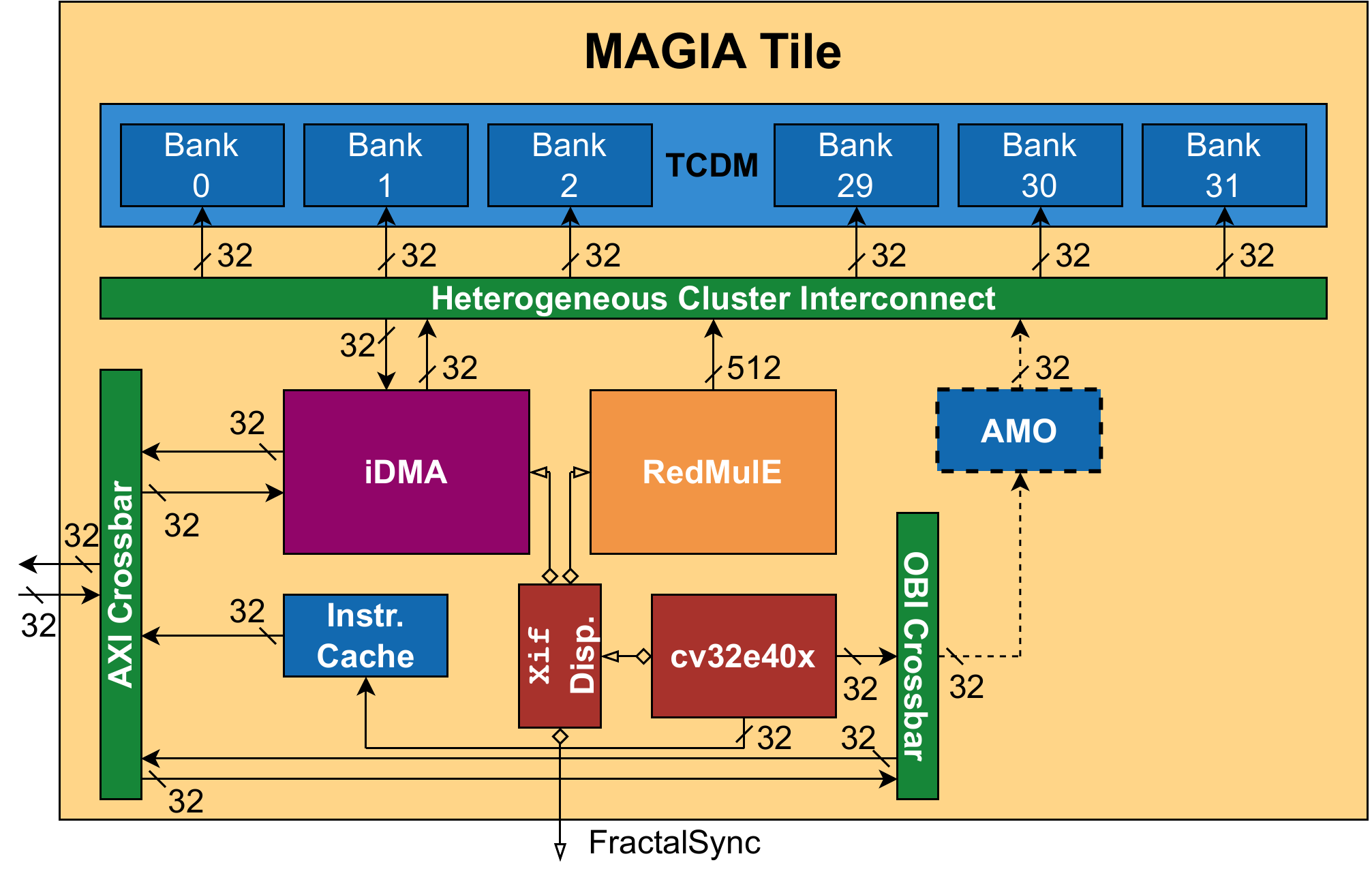}
  \caption{Architecture of MAGIA tile featuring 32 TCDM banks, dual-channel 32-bit DMA, 24$\times$8 semi-systolic GEMM accelerator and an AMO management module.}
  \Description{View of the MAGIA tile representing the GEMM accelerator, DMA, L1 SPM, controll core (cv32e40x), instruction cache, instruction dispatcher, optional AMO hardware module and the interconnect.}
  \label{fig:tile}
\end{figure}

At the heart of \textit{MAGIA}\footnote{\url{https://github.com/pulp-platform/MAGIA}} - Mesh Architecture for Generative Intelligence Acceleration - lies the tile containing a GEMM accelerator, a DMA engine, a multi-banked L1 SPM and a lightweight control core. Figure \ref{fig:tile} shows a particular configuration of the tile. The L1 features 32 interleaved memory banks that compose the Tightly-Coupled Data Memory (TCDM). Each bank is of 32 KiB and 32-bit wide, for a total of 1MiB per tile. The L1 can be accessed through a single-cycle latency Heterogeneous Cluster Interconnect (HCI)\footnote{\url{https://hwpe-doc.rtfd.io}}. Each tile has access to the global L2  and to a subset of other tiles' L1, accessing the latter via one-sided DMA reads/writes. Inter-tile and global communication is carried out through a 2-channel 32-bit AXI4 crossbar (Xbar). External tiles and the core access the L1 through an OpenHW's OpenBus Interface (OBI) Xbar and an AMO hardware module.

Each tile is controlled by a cv32e40x\footnote{\url{https://github.com/pulp-platform/cv32e40x}}\cite{gautschi2017near} core coupled with 16KiB of instruction cache. The system has been extended with custom instructions to program and control the iDMA, RedMulE, and synchronization hardware (\textit{FractalSync}). These instructions are implemented using OpenHW's eXtension Interface (\texttt{Xif})\footnote{\url{https://docs.openhwgroup.org/projects/openhw-group-core-v-xif}}. The latter allows us to extend the ISA of the core without modifying its micro-architecture. A dedicated module dispatches instructions not meant for the core to the appropriate module based on the instruction signature (OPCODE and FUNC3).

MatMul kernels are offloaded to RedMulE\footnote{\url{https://github.com/pulp-platform/redmule}}\cite{tortorella2023redmule}: an open-source accelerator design for GEMM-Ops. The configuration we integrate in the tile is characterized by a 24$\times$8 semi-systolic array and 512 bits of L1 interface bandwidth.

To support fast communication and enable its overlapping with computation, we integrate an open-source DMA engine: iDMA\footnote{\url{https://github.com/pulp-platform/iDMA}}\cite{benz2023high}. The particular instance we use has two separate 32-bit transfer channels, one for moving data from outside the tile to the L1 and one for traffic in the opposite direction. To program the iDMA we add a dedicated hardware module that interfaces between the core and the iDMA.

\subsection{NoC}
\begin{figure}[t]
  \centering
  \includegraphics[width=0.8\linewidth]{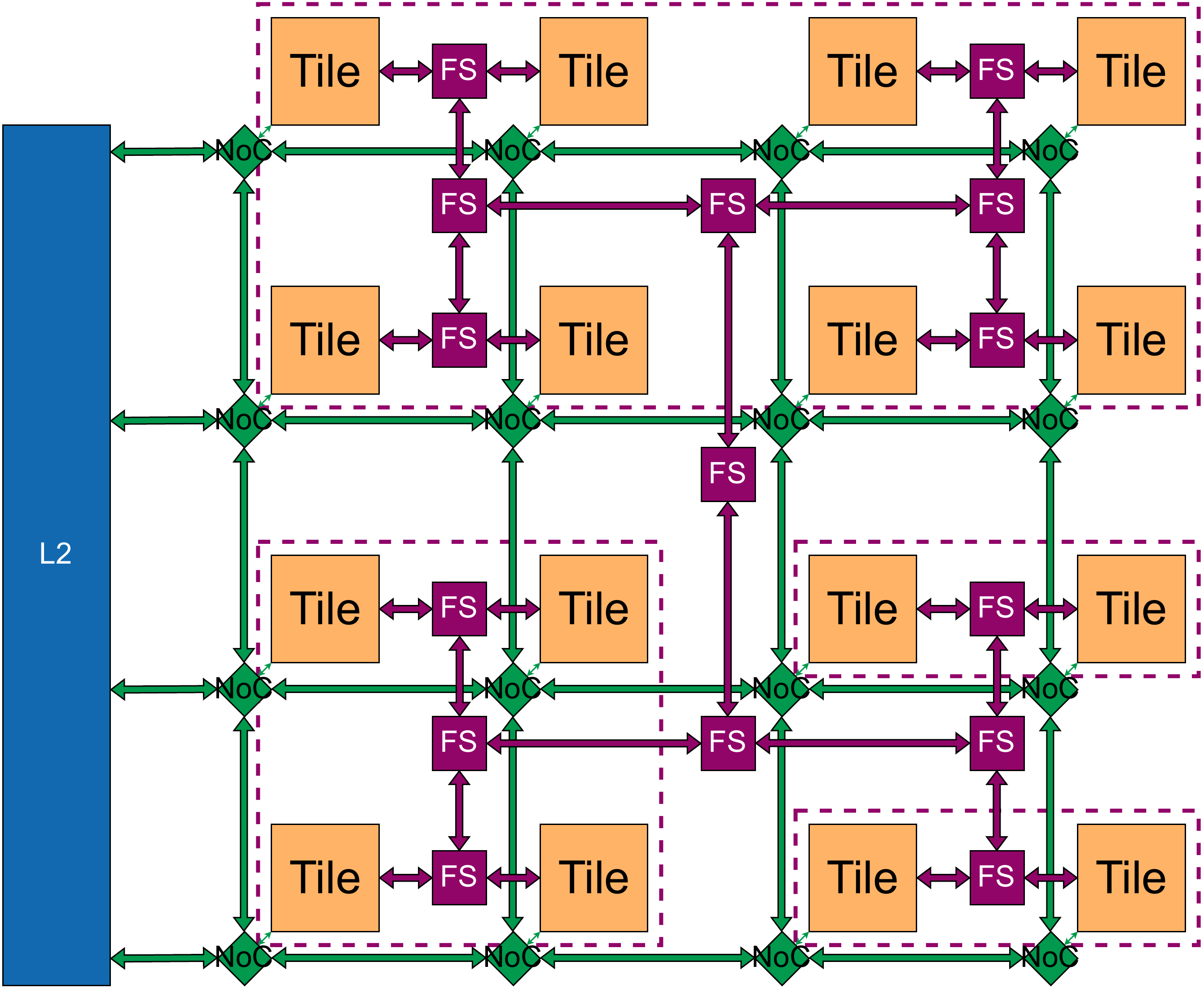}
  \caption{Architecture of a 4$\times$4 configuration of MAGIA featuring the L2, the NoC and the H-tree FractalSync (FS) synchronization network. Purple dashed lines indicate synchronization domains.}
  \Description{Overview of the MAGIA representing the tiles, the NoC and the synchronization network.}
  \label{fig:mesh}
\end{figure}

Replicating the \textit{MAGIA} tile, we scale up to a homogeneous mesh of compute tiles. As shown in Figure \ref{fig:mesh}, we choose a two-dimensional (2D) mesh topology: the top-level instantiates the \textit{MAGIA} tiles and the interconnect and provides a number of interfaces towards the L2 memory equal to the number of rows of the mesh.

To guarantee both the bandwidth for efficient data movement and the scalability of the system, we instantiate the open-source \textit{FlooNoC}\footnote{\url{https://github.com/pulp-platform/FlooNoC}} \cite{fischer2025floonoc} Network-on-Chip. Reflecting the top-level architecture, we instantiate a 2D XY mesh that features 32-bit physical links. The conversion between the AXI4 protocol, used by the compute tiles, and the network-level protocol is performed by Network Interfaces (NIs) between each tile and the near router.

The choice of a single physical link with no virtual channels can be justified when synchronization between tiles is managed through a dedicated network (e.g. \textit{FractalSync}). If that is the case, the traffic flowing within the NoC is related only to the movement of data and instructions between the L2 memory and the tiles.

\section{FractalSync} \label{sec:fsync}

\begin{figure}[t]
  \centering
  \includegraphics[width=\linewidth]{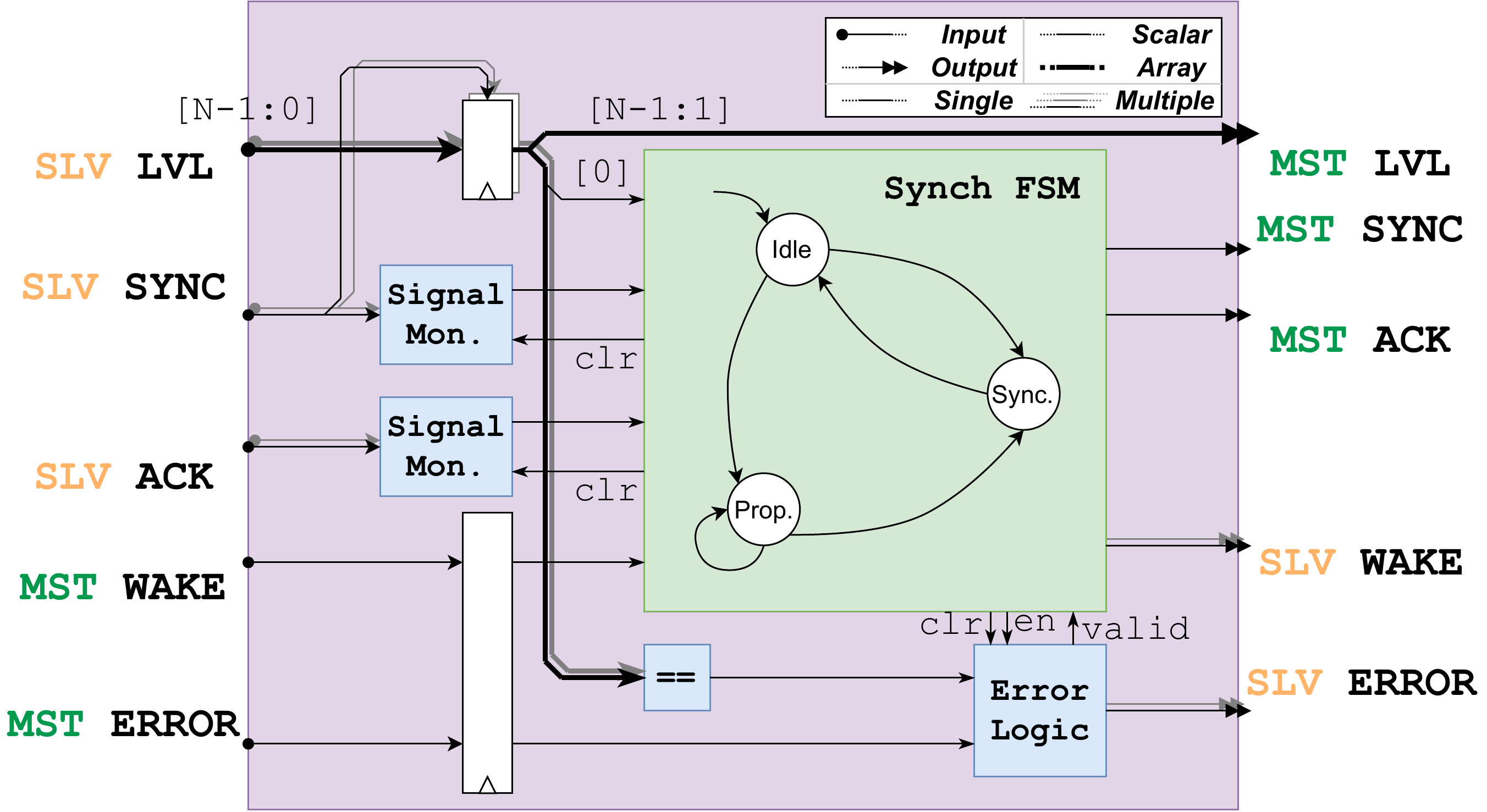}
  \caption{Overview of the micro-architecture of the FractalSync module.}
  \Description{micro-architecture of the FractalSync module indicating the main FSM, peripheral logic and datapaths.}
  \label{fig:fsync}
\end{figure}

\subsection{Synchronization Tree}
The barrier synchronization mechanism is based on a divide-and-conquer strategy. We first tackle the synchronization of two neighboring tiles with a dedicated fast and lightweight hardware module - \textit{FractalSync}\footnote{\url{https://github.com/VictorIsachi/fractal_sync/tree/fsync_v1}}. The mesh of $k$$\times$$k$ PEs is divided into $k^2/2$ subsets, each containing 2 distinct neighboring PEs. Each subset is synchronized by a dedicated \textit{FractalSync}. We now have a set of $k^2/2$ synchronization modules that need to be synchronized themselves. We follow a similar approach, dividing the $k^2/2$ \textit{FractalSync}s into $k^2/4$ subsets, each containing 2 distinct neighboring synchronization modules, and synchronizing them. We proceed recursively until we have only $2$ \textit{FractalSync}s that need to be synchronized. This is done by the root of the synchronization tree that is formed by this process. The recursive nature of this scheme naturally maps to an H-tree, as can be seen in Figure \ref{fig:mesh}. This topology is well known to have desirable properties, including its optimality in covering a 2D surface~\cite{leiserson1980area}.

\subsection{Programmability} 
\label{subsec:programmability}
\textit{FractalSync} is capable of synchronizing not only the entire mesh, but also specific subsets of it. Given a node of the synchronization tree, all PEs having that node as root can synchronize separate of other PEs in the mesh. We call a set of PEs that synchronize together a \textit{synchronization domain}. For example, the mesh of Figure \ref{fig:mesh} depicts the 8 upmost tiles forming a group that is synchronized separately from the rest of the mesh. The 4 leftmost remaining tiles form a different synchronization domain, and the remaining tiles form another 2 synchronization domains.

From the programming point of view, synchronization can be implemented with a single instruction: \texttt{fsync(level)}. The argument \texttt{level} indicates the node of the synchronization tree the tile should synchronize at. This programming model is implemented by extending the ISA of \textit{MAGIA}'s control core with a dedicated instruction via \texttt{Xif}. Additionally, we need a specialized instruction decoder that interfaces between the synchronization tree and \texttt{Xif} dispatcher to forward synchronization requests/responses.

\subsection{Micro-architecture}
A synchronization request can be initiated in one of the two slave ports of the module via the \textit{req} channel. This channel is composed by \textit{sync} and \textit{lvl}. The former is a signal that indicates a synchronization request, the latter specifies the level of the tree. Synchronization is served through the \textit{rsp} channel, with \textit{wake} and \textit{error} indicating occurred synchronization and detected error (e.g. mismatch in synchronization level of neighboring tiles) respectively. The \textit{wake} signal is asserted until the module detects the assertion of all \textit{ack} signals. This indicates that all requesting tiles have been synchronized and the module should be free to accept new synchronization requests.

We use one-hot encoding to represent levels of the synchronization tree. This allows us to effectively consider a synchronization request as a 2-wire transaction since only 2 wires (i.e. \textit{sync} and one wire of \textit{lvl}) will be concurrently asserted during a transaction. We can afford to use one-hot encoding since we only need $2\log_2(k)$ wires to encode levels in a $k$$\times$$k$ mesh. Furthermore, each time we climb to the next level of the tree, we can discard a wire since we know that synchronization will not occur at the current level.

The micro-architecture of \textit{FractalSync} is summarized in Figure \ref{fig:fsync}. Note that \textit{lvl} is sampled only when \textit{sync} is asserted and that the least significant bit (lsb) of the sample is used by the main Finite State Machine (FSM) - Synch FSM. The latter manages slave synchronization and master request propagation. Based on \textit{lvl}'s lsb, it determines if synchronization must occur at the current level of the tree. The rest of \textit{lvl}'s bits are propagated to the next level of the tree.

Figure \ref{fig:fsync} shows that the datapath is duplicated for slave ports. The same \textit{wake} and \textit{error} signals are propagated to both. The \textit{lvl} signal of only one slave port is used to determine whether synchronization should be propagated, and the level encoding, since a mismatch is an error detected and handled by dedicated logic. \textit{sync} and \textit{ack} are monitored by sequential logic that reports when both ports have asserted, not necessarily in the same cycle, the respective signals.

\section{Experimental Evaluation}
\label{sec:results}

\subsection{Performance} \label{sec:speedup}
We implemented the \textit{MAGIA} architecture and \textit{FractalSync} in synthesizable SystemVerilog RTL.
We carried out all performance evaluations with cycle-accurate RTL simulations of the full mesh architecture in various configurations, spanning from 2$\times$2 through 16$\times$16 tiles, plus a simple case with only two tiles (\textit{Neighbour}).
To accurately measure the impact of NoC communication in baseline AMO synchronization schemes, we mitigate the noise introduced by cache misses by pre-heating instruction caches with appropriate synchronization instructions. 

Let $P = \{p_1, \dots, p_N\}$ be a set of $N$ PEs and let $R=\{r_1, \dots, r_N\}$ and $F=\{f_1, \dots, f_N\}$ be the sets, respectively, of associated cycles in which PEs request synchronization and in which they execute the instruction following synchronization. Then, \textit{synchronization overhead}, $\hat{S}$ with unit cycles $[c]$, is defined as $\hat{S} := \max(F) - \max(R)$. The latter is the main metric we use to evaluate performance of synchronization schemes.

The baseline we evaluate \textit{FractalSync} against is composed of two synchronization schemes: \textit{Na\"ive} and \textit{XY}.
Na\"ive designates a single tile responsible for accepting synchronization requests and dispatching synchronization responses.
The Na\"ive scheme is very simple, which means it has very little associated instruction overhead; however, its quadratic scaling limits its usefulness to small meshes. XY, on the other hand, breaks down the problem from one 2D synchronization to two 1D synchronizations by first synchronizing along one direction of the mesh and then the other one.
The XY scheme, compared to Na\"ive, adds significant instruction overhead; however, its scaling is linear, which can be proven to be optimal.

The \textit{FractalSync}-based approach relies on hardware synchronization trees organized as shown in Figure~\ref{fig:mesh}.
In the \textit{FractalSync} case, each tile's core simply calls the implemented \texttt{fsync()} function discussed in Subsection~\ref{subsec:programmability} to perform synchronization at the root of the synchronization tree.

Note that for large meshes the distance between nodes in the synchronization tree can exceed the distance between neighboring NoC nodes.
For a fair comparison, together with the native \textit{FractalSync} case, in our experiments we also consider the effect of adding pipeline stages to break these longer connections into segments that do not exceed in length the distance between two neighboring NoC nodes, as an additional \textit{FractalSync+Pipeline} case.

\begin{table}[t]
    \centering
    \caption{Performance of synchronization schemes. FSync stands for native FractalSync, and FSync+P for FractalSync+Pipeline. For each mesh configuration we highlight the best-performing NoC-based scheme.}
\resizebox{\columnwidth}{!}{
    \begin{tabular}{|c|c|c|c|c|c|}
    \hline
     
    \textbf{Mesh}    & \textbf{FSync}   & \textbf{FSync+P} & \textbf{Na\"ive} & \textbf{XY}    &  \textbf{Speedup} \\ 
    \textbf{config.} & \textbf{($c$)}   & \textbf{($c$)}   & \textbf{($c$)}   & \textbf{($c$)} &  \textbf{Fsync+P vs.} \\ 
                     &                  &                  &                  &                &  \textbf{Best AMO} \\ \hline
    \multicolumn{6}{|c|}{\textbf{--- 16KiB instruction cache, cache pre-heating ---}} \\ \hline
    $Neighbor$   & $4$  & $4$  & $\textbf{79}$    & $\textbf{79}$   & $\textbf{19}\times$ \\ \hline
    2$\times$2   & $6$  & $6$  & $\textbf{119}$   & $219$  & $\textbf{19}\times$ \\ \hline
    4$\times$4   & $10$ & $10$ & $512$            & $\textbf{347}$  & $\textbf{34}\times$ \\ \hline
    8$\times$8   & $14$ & $18$ & $2488$           & $\textbf{614}$  & $\textbf{34}\times$ \\ \hline
    16$\times$16 & $18$ & $34$ & $13961$          & $\textbf{1462}$ & $\textbf{43}\times$ \\ \hline
    \end{tabular}
    \label{tab:sync_perf}
}
\end{table}

Table \ref{tab:sync_perf} reports the performance of all the synchronization schemes we discussed.
We report the speed-up as the ratio of the fastest NoC-based synchronization algorithm and the \textit{FractalSync+Pipeline} scheme. What we be observed in these results is: \textit{(i)} \textit{FractalSync} scales better than solutions based on atomic memory operations, this can be seen in the fact that speed-up increases as the mesh scales out. \textit{(ii)} Even for small meshes and neighboring tiles the speed-up is significant ($19\times$), reaching a maximum of $43\times$ for the largest mesh configuration we tested. \textit{(iii)} As expected, Na\"ive beats XY for small meshes but quickly becomes worse due to its suboptimal scaling.

\subsection{Area evaluation} \label{subsec:logical_synchesis}

\begin{figure}[t]
    \centering
    \includegraphics[width=\linewidth]{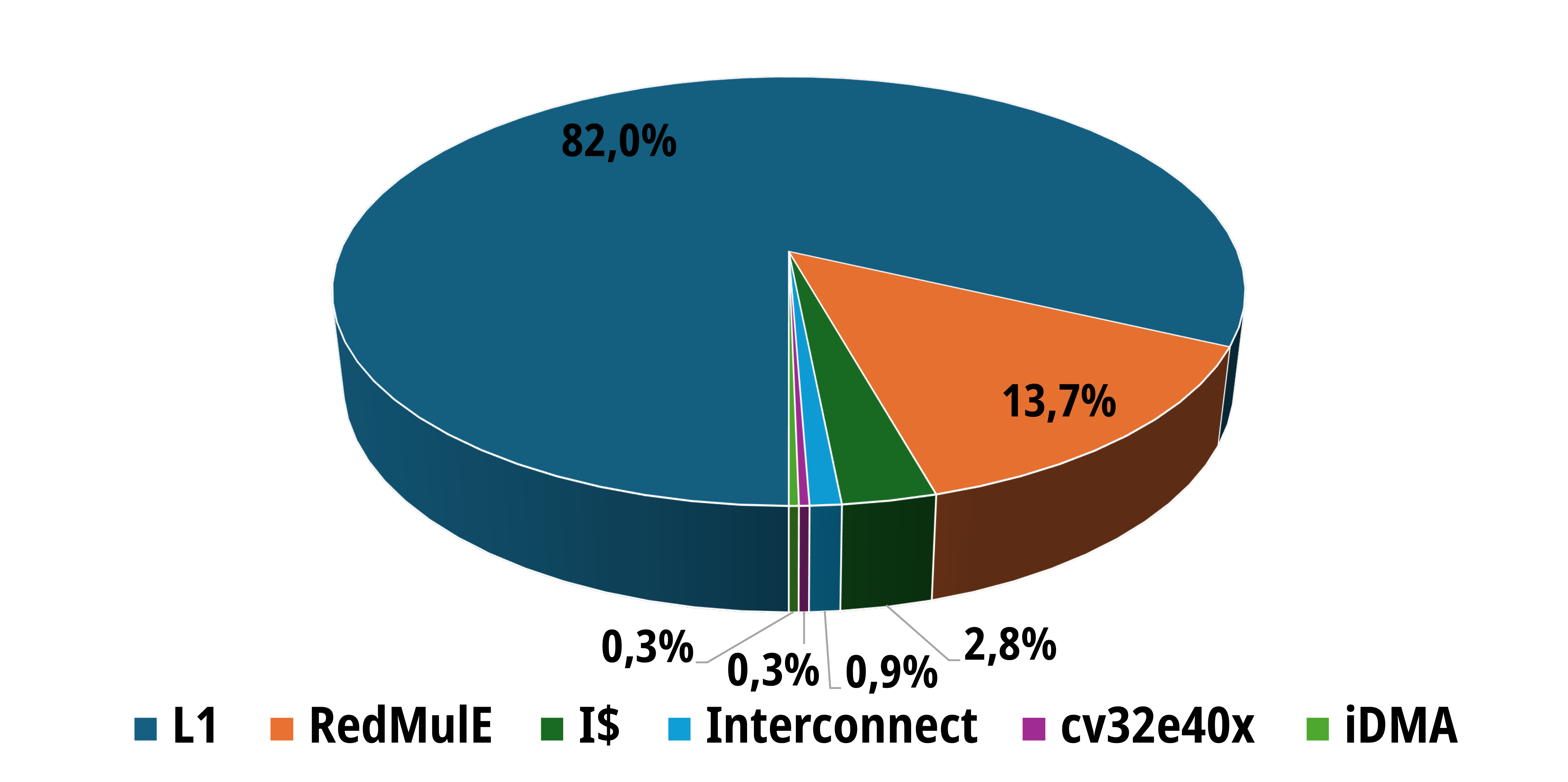}
    \caption{MAGIA tile area breakdown.}
    \Description{MAGIA tile area breakdown among L1, RedMulE, instruction cache, interconnect, cv32e40x and iDMA.}
    \label{fig:tile_area}
\end{figure}

We performed the logic synthesis of the \textit{MAGIA} tile, the NoC and the \textit{FractalSync} module in \textsc{GlobalFoundries}' 12nm FinFet technology using Design Compiler by \textsc{Synopsys}. We synthesized our design targeting 1 GHz clock frequency in SSPG corner at -40\degree C.
We performed the logic synthesis of the \textit{MAGIA} tile considering two configurations: \textit{(i)} one that supports the synchronization schemes based on the NoC and AMO instructions and \textit{(ii)} one that adds \textit{FractalSync} on top. The results we obtain are: 1.5816 $mm^ 2$ for configuration \textit{(i)} and 1.5814 $mm^ 2$ for \textit{(ii)}. The slight area decrease can be attributed to the noise introduced by the logic synthesis tool. These results indicate that the two configurations are comparable in terms of area and that the synchronization mechanism based on \textit{FractalSync} does not have a relevant impact on the area of the tile. We analyzed the area distribution within the tile with support for both AMO and \textit{FractalSync} and reported it in Figure \ref{fig:tile_area}. The contributions of both the AMO module and \textit{FractalSync} is < 0.03\% and for this reason does not appear in the breakdown.

We modeled the total area of the full system exploiting the synthesis results of the single elements composing it. The estimates we obtain consider the fact that the top-level integrates a $k$$\times$$k$ NoC, $k^2$ \textit{MAGIA} tiles, and $k^2-1$ \textit{FractalSync} (FS) modules. Even without considering the contribution of the memory banks within the tiles to the total area of \textit{MAGIA}, the maximum overheads given by the NoC and the synchronization network are, respectively, 1.7\% and 0.007\% while more than 98\% of the area is dedicated to the computation and communication logic.

\section{Conclusion} \label{sec:conclusion}
In this paper we introduced \textit{FractalSync}, a fast and lightweight barrier synchronization scheme. We also introduced \textit{MAGIA}, a massively parallel tile-based AI accelerator. We integrated \textit{FractalSync} in \textit{MAGIA} and demonstrated that this addition speeds up synchronization by up to 43$\times$ in our setup. Furthermore, our findings indicate that the proposed synchronization scheme scales significantly better than the baseline NoC-based SW synchronization. We studied the area efficiency of our solution showing that it adds negligible overhead while being able to close the 1GHz target timing of the host system.

\begin{acks}
This work was supported by NextGenerationEU PNRR CN HPC, Spoke 1 Future HPC.
\end{acks}

\bibliographystyle{ACM-Reference-Format}
\bibliography{sample-base}

\end{document}